\documentclass[showpacs,aps,prb,twocolumn,groupedaddress]{revtex4}
\usepackage{graphicx}

\begin{document}


\title{Electronic correlation in the infrared optical properties of the quasi two dimensional $\kappa$-type BEDT-TTF dimer system.}


\author{T. Sasaki}
\author{I. Ito}
\author{N. Yoneyama}
\author{N. Kobayashi}
\affiliation{Institute for Materials Research, Tohoku University, Katahira 2-1-1, Sendai 980-8577, Japan}
\author{N. Hanasaki}
\author{H. Tajima}
\affiliation{Institute for Solid State Physics, University of Tokyo, Kashiwanoha 5-1-5, Kashiwa 277-8581, Japan}
\author{T. Ito}
\affiliation{JAIST Hokuriku, Asahidai 1-1, Tatsunokuchi, Ishikawa 923-1292, Japan}
\author{Y. Iwasa}
\affiliation{Institute for Materials Research, Tohoku University, Katahira 2-1-1, Sendai 980-8577, Japan}
\affiliation{CREST, Japan Science and technology Corporation, Kawaguchi 332-0012, Japan}

\date{\today}

\begin{abstract}

The polarized optical reflectance spectra of the quasi two dimensional organic correlated electron system $\kappa$-(BEDT-TTF)$_{2}$Cu[N(CN)$_{2}$]$Y$, $Y =$ Br and Cl are measured in the infrared region.  The former shows the superconductivity at $T_{\rm c} \simeq$ 11.6~K and the latter does the antiferromagnetic insulator transition at $T_{\rm N} \simeq$ 28~K.  
Both the specific molecular vibration mode $\nu_{3}(a_{g})$ of the BEDT-TTF molecule and the optical conductivity hump in the mid-infrared region change correlatively at $T^{*} \simeq$ 38~K of $\kappa$-(BEDT-TTF)$_{2}$Cu[N(CN)$_{2}$]Br, although no indication of $T^{*}$ but the insulating behaviour below $T_{\rm ins} \simeq$ 50-60 K are found in $\kappa$-(BEDT-TTF)$_{2}$Cu[N(CN)$_{2}$]Cl.  
The results suggest that the electron-molecular vibration coupling on the $\nu_{3}(a_{g})$ mode becomes weak due to the enhancement of the itinerant nature of the carriers on the dimer of the BEDT-TTF molecules below $T^{*}$, while it does strong below $T_{\rm ins}$ because of the localized carriers on the dimer.  
These changes are in agreement with the reduction and the enhancement of the mid-infrared conductivity hump below $T^{*}$ and $T_{\rm ins}$, respectively, which originates from the transitions between the upper and lower Mott-Hubbard bands. 
The present observations demonstrate that two different metallic states of $\kappa$-(BEDT-TTF)$_{2}$Cu[N(CN)$_{2}$]Br are regarded as {\it a correlated good metal} below $T^{*}$ including the superconducting state and {\it a half filling bad metal} above $T^{*}$.  
In contrast the insulating state of $\kappa$-(BEDT-TTF)$_{2}$Cu[N(CN)$_{2}$]Cl below $T_{\rm ins}$ is the Mott insulator.

\end{abstract}


\maketitle


\section{Introduction}

Organic charge transfer salts based on the donor molecule bis(ethylenedithio)-tetrathiafulvalene, abbreviated BEDT-TTF, are characterized by their quasi-two dimensional (Q2D) electronic properties which originate in the layer structure of (BEDT-TTF)$_{2}X$ crystals.\cite{Ishiguro,Lang1}  
The BEDT-TTF molecules have different possible packing patterns in the conducting layer.  
The patterns are denoted by the different Greek letters.  
Among them, $\kappa$-type has a unit of dimer consisting of two BEDT-TTF molecules.  
Since each dimer has one hole (three electrons) because of charge transfer to the anions, the conduction band can be considered to be effectively half-filling.  
Therefore, $\kappa$-(BEDT-TTF)$_{2}$$X$ with $X =$ Cu(NCS)$_{2}$ and Cu[N(CN)$_{2}$]$Y$ ($Y =$ Br and Cl) have attracted considerable attention from the point of view of the strongly correlated Q2D electron system.\cite{Kanoda1,Kanoda2,McKenzie2,Singleton1}  
The unconventional metallic, antiferromagnetic insulator and superconducting phases appear next to one another in the phase diagram.\cite{Kanoda1,Kanoda2,Sasaki1}  
The transitions among these phases are controlled by the applied pressure which must change the conduction band width $W$ with respect to the effective Coulomb repulsion $U$ between two electrons on a dimer.  
Thus the $\kappa$-(BEDT-TTF)$_{2}$$X$ family has been considered to be the band width controlled Mott system in comparison with the filling controlled one in the inorganic perovskites like as High-$T_{\rm c}$ copper oxides.\cite{McKenzie1,Imada}  

Recently anomalies in the metallic state of the superconducting salts at $T^{*} \simeq$ 38 K for $X =$ Cu[N(CN)$_{2}$]Br and 50 K for $X =$ Cu(NCS)$_{2}$ have been reported in several physical properties; the cusp of the spin-lattice relaxation rate $(T_{1}T)^{-1}$ appears in $^{13}$C-NMR,\cite{Mayaffre1,Kawamoto1,Soto1} the sound velocity takes a pronounced minimum,\cite{Frikach,Shimizu,Fournier} the thermal expansion coefficient indicates a reminiscent of second-order phase transition,\cite{Mueller} and the spin susceptibility and the resistivity show a change of the anisotropy.\cite{Sasaki1}  
The $T^{*}$ line seems to divide the metallic phase into unconventional metal with large antiferromagnetic spin fluctuation at $T > T^{*}$ and a metal with a possibility of the fluctuated density wave formation at $T < T^{*}$.\cite{Sasaki1}  
The line has been found to be terminated at the critical end point of the first order Mott metal-insulator transition line.\cite{Lefebvre,Limelette,Kagawa}  
The nature of the $T^{*}$ line (transition or crossover) and these unconventional metallic regions, however, has not been understood well.  
It is important particularly to investigate the characteristics in the metallic state at $T_{\rm c} < T < T^{*}$ because it must be closely related to the mechanism and the symmetry of the superconductivity which has been controversially discussed so far.\cite{Lang1,Izawa}  
In this point the anticorrelation of the sings of the uniaxial-pressure coefficients at $T_{\rm c}$ and $T^{*}$ has been pointed out from a thermodynamic analysis.\cite{Lang2}
It is noted again that such $T^{*}$ anomalies and the metallic state below $T^{*}$ appear in only the superconducting salts and those have not been found in the no-superconducting salts, for example, the antiferromagnetic (AF) Mott insulator $\kappa$-(BEDT-TTF)$_{2}$Cu[N(CN)$_{2}$]Cl, which has the AF transition at $T_{\rm N}$ = 28~K \cite{Miyagawa} and the insulator-semiconductor like transition at $T_{\rm ins} \simeq$ 50-60~K.\cite{Limelette,Ito}

Optical conductivity measurements are powerful tool for investigating the charge dynamics, electron correlation, and electron-phonon (molecular vibration) coupling in the correlated electron system like the low dimensional organic conductors.\cite{Jacobsen}  
In this paper, we report the systematic infrared (IR) reflectance investigation of $\kappa$-(BEDT-TTF)$_{2}$Cu[N(CN)$_{2}$]$Y$ ($Y =$ Br and Cl).  
We focus on the change of a specific molecular vibration mode $\nu_{3}(a_{g})$ and the broad conductivity peak in the mid-IR region of the superconducting $\kappa$-(BEDT-TTF)$_{2}$Cu[N(CN)$_{2}$]Br and the AF-Mott insulator $\kappa$-(BEDT-TTF)$_{2}$Cu[N(CN)$_{2}$]Cl. 
The symmetric $\nu_{3}(a_{g})$ mode becomes visible in the IR spectra due to the large electron-molecular vibration (EMV) coupling.\cite{Jacobsen,Rice,Rice2}  
Thus the mode is sensitive to the electronic state on the dimer molecules.  
The conductivity peak at mid-IR has been resulted in the inter-band transitions based on the tight binding band calculation.\cite{Tajima,Kuroda,Eldridge1,Kornelsen1,Tamura,Kornelsen2}   
In the case of strongly correlated electron system with half-filling, however, a broad band has been expected to appear at the similar mid-IR region corresponding to the energy of $U$ which is attributed to the transitions between the lower and upper Hubbard bands.\cite{McKenzie2,Rozenberg,Georges,Merino1}  
We discuss the change of the electronic state and the correlation effect of $\kappa$-(BEDT-TTF)$_{2}$$X$ by probing such optical properties.  

\section{Experiments}

High quality single crystals of $\kappa$-(BEDT-TTF)$_{2}$Cu[N(CN)$_{2}$]$Y$ with $Y =$ Br and Cl were grown by the standard electrochemical oxidation method.  
The crystals have a slightly distorted hexagonal and diamond shaped shiny surface containing the $c$ and $a$-axes.  
The polarized reflectance spectra were measured on the $c$-$a$ plane along $E \parallel a$-axis and $E \parallel c$-axis with a Fourier transform microscope-spectrophotometer (JASCO FT/IR-620 and MICRO-20).  
The mid-IR (600 - 6000 cm$^{-1}$) region was measured by use of a mercury-cadmium-telluride (MCT) detector at 77 K and a KBr beamsplitter.  
The sample was fixed by the conductive carbon paste on the microgoniometer which was placed at the cold head of the helium flow type refrigerator (Oxford CF1104).  
The temperature was monitored and controlled at 10 - 300 K by the gold/iron-Chromel thermocouple and diode thermometers.  
We gave careful consideration of less stress and good thermal contact to the crystals in the sample setting.  
The reflectivity was obtained by comparison with a gold mirror at room temperature.  
The optical conductivity was calculated by a Kramers-Kronig analysis of the reflectivity.  
For this analysis, the reflectivity data was extrapolated to low and high frequencies.  
The conductivity in the measured frequency range was found to be insensitive to the small difference of the extrapolations.  
The Raman spectra were measured by using a JASCO NRS-1000 with a backscattering geometry.  
We used a He-Ne laser with the wavelength at 632.8 nm.

\section{Results and Discussion}

\begin{figure}
\includegraphics[viewport=1cm 3cm 20cm 28cm,clip,width=0.9\linewidth]{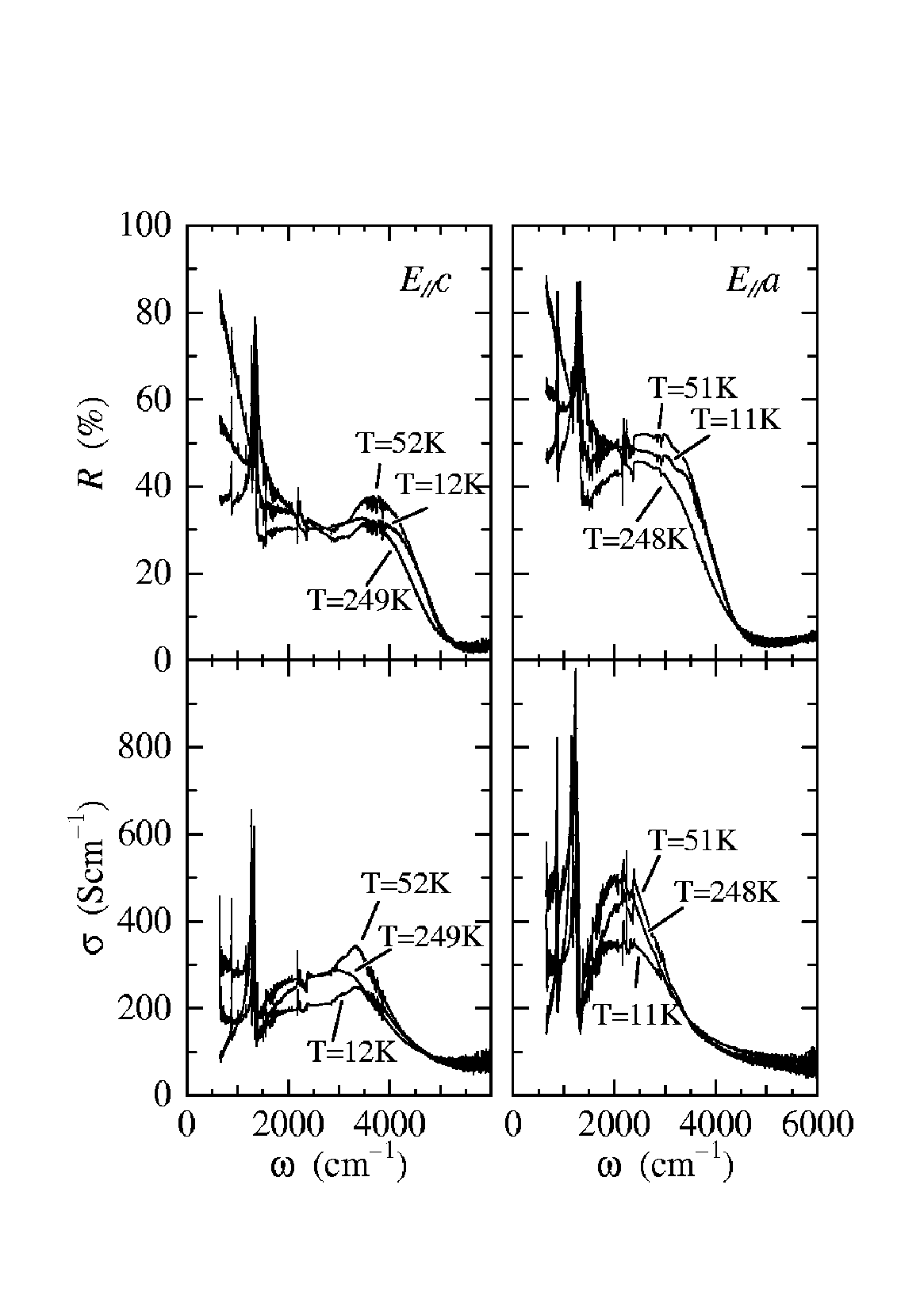}
\caption{The reflectivity (upper panels) and the conductivity spectra (lower panels) of $\kappa$-(BEDT-TTF)$_{2}$Cu[N(CN)$_{2}$]Br for $E \parallel$ $c$-axis (left panels) and $E \parallel$ $a$-axis (right panels), respectively.  }
\end{figure}

Figure 1 shows the reflectivity (upper panels) and the optical conductivity spectra (lower panels) calculated by the Kramers-Kronig transformation for $E \parallel c$-axis (left side panels) and $E \parallel a$-axis (right side panels) of $\kappa$-(BEDT-TTF)$_{2}$Cu[N(CN)$_{2}$]Br which shows superconductivity at 11.6~K.\cite{Kini}  
The data taken at three temperatures are selected to be shown in the figure for clarity.  
The features of both the reflectance and the conductivity spectra reproduce well the reported results.\cite{Eldridge1,Vlasova,Griesshaber}  
Characteristic features of the spectra are; 1) Drude peak appears only at low temperature, 2) the broad mid-IR conductivity peak exists at 3300 cm$^{-1}$ for $E \parallel c$-axis and 2200 cm$^{-1}$ for $E \parallel a$-axis, and 3) the molecular vibrational modes of BEDT-TTF are superimposed on the electronic background.  
On the Drude behaviour, we are not involved in the detail of its temperature dependence because the present experiments are limited in the measurable frequency range down to 600 cm$^{-1}$.  
The tendency of low frequency region, however, is in good agreement with the previous work.\cite{Eldridge1}  
As the intensity of the far-IR conductivity has grown with decreasing temperature, the strength of the mid-IR conductivity has decreased.  
The temperature dependence of the mid-IR conductivity is discussed latter in connection with a characteristic temperature $T^{*} =$ 38~K and the molecular vibrational modes.  

\begin{figure}
\includegraphics[viewport=1cm 3cm 20cm 28cm,clip,width=0.9\linewidth]{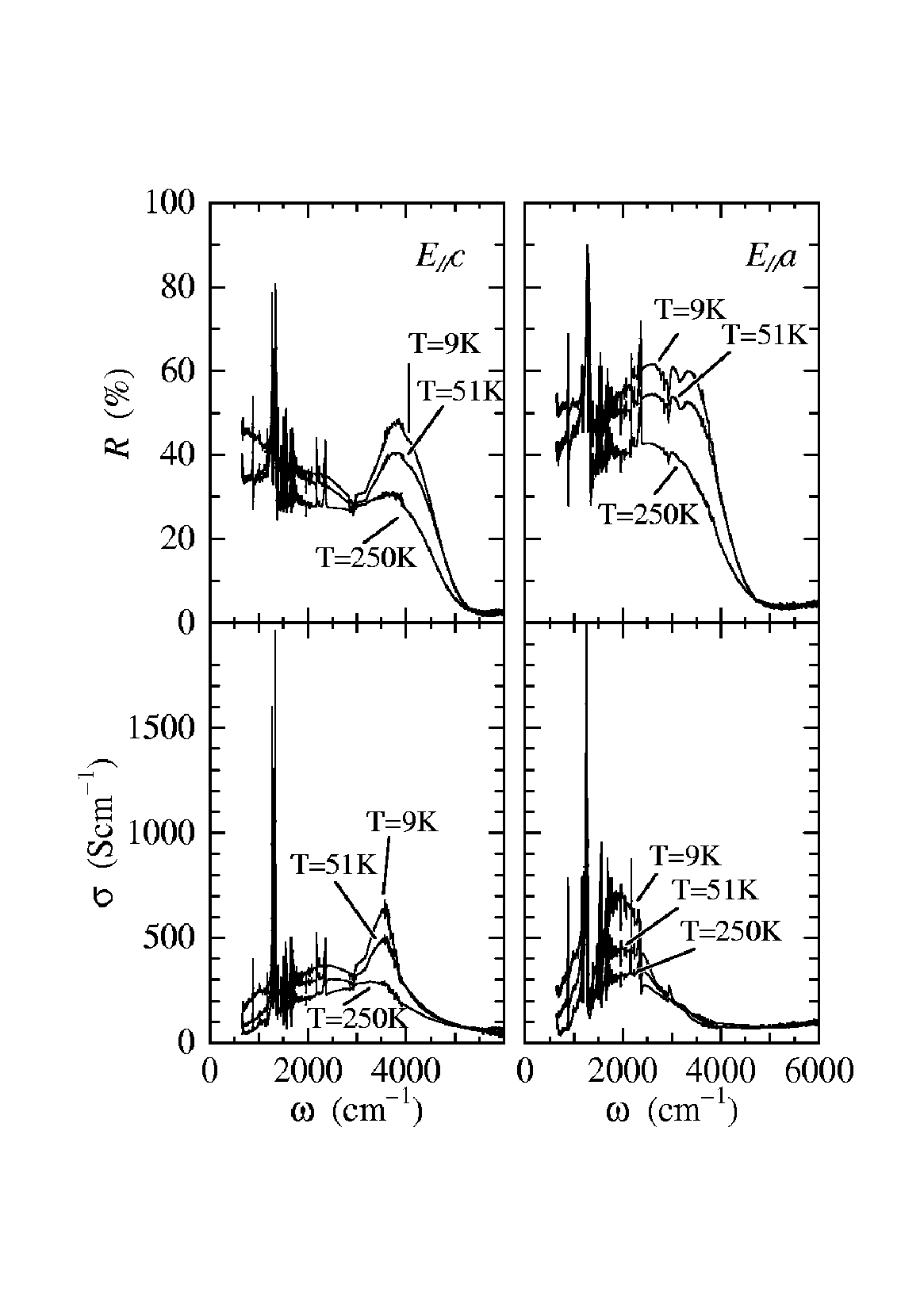}
\caption{The reflectivity (upper panels) and the conductivity spectra (lower panels) of $\kappa$-(BEDT-TTF)$_{2}$Cu[N(CN)$_{2}$]Cl for $E \parallel$ $c$-axis (left panels) and $E \parallel$ $a$-axis (right panels), respectively.  }
\end{figure}

Figure 2 shows the reflectivity (upper panels) and the optical conductivity spectra (lower panels) calculated by the Kramers-Kronig transformation for $E \parallel c$-axis (left side panels) and $E \parallel a$-axis (right side panels) of $\kappa$-(BEDT-TTF)$_{2}$Cu[N(CN)$_{2}$]Cl which shows antiferromagnetic insulator transition at 27~K.\cite{Miyagawa}  
The features of both the reflectance and the conductivity spectra also reproduce well the results which have been reported.\cite{Kornelsen1,Vlasova}  
Below about 50-60~K, in contrast to the superconducting salt, the conductivity at low frequency range decreases and the mid-IR conductivity and a vibrational mode $\nu_{3}(a_{g})$ are strongly enhanced, which means that the mid-IR band transitions are becoming dominant at lower temperatures.  

\begin{figure}
\includegraphics[viewport=1cm 4.5cm 20cm 26cm,clip,width=0.9\linewidth]{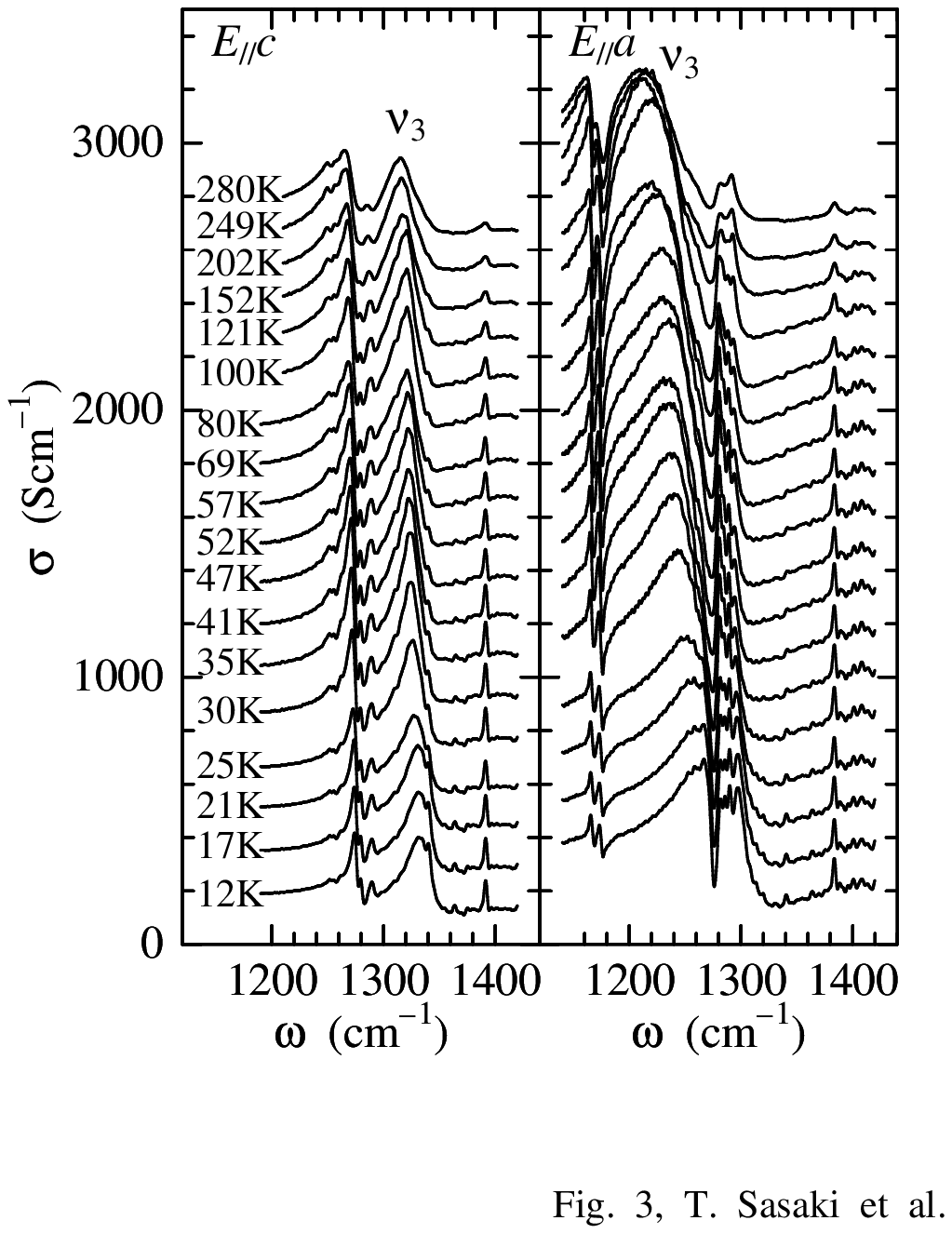}
\caption{The conductivity spectra near 1300 cm$^{-1}$ of $\kappa$-(BEDT-TTF)$_{2}$Cu[N(CN)$_{2}$]Br for $E \parallel$ $c$-axis (left) and $E \parallel$ $a$-axis (right).  The spectra are plotted with shifting the conductivity by 150 Scm$^{-1}$ each.  }
\end{figure}

\begin{figure}
\includegraphics[viewport=1cm 4.5cm 20cm 26cm,clip,width=0.9\linewidth]{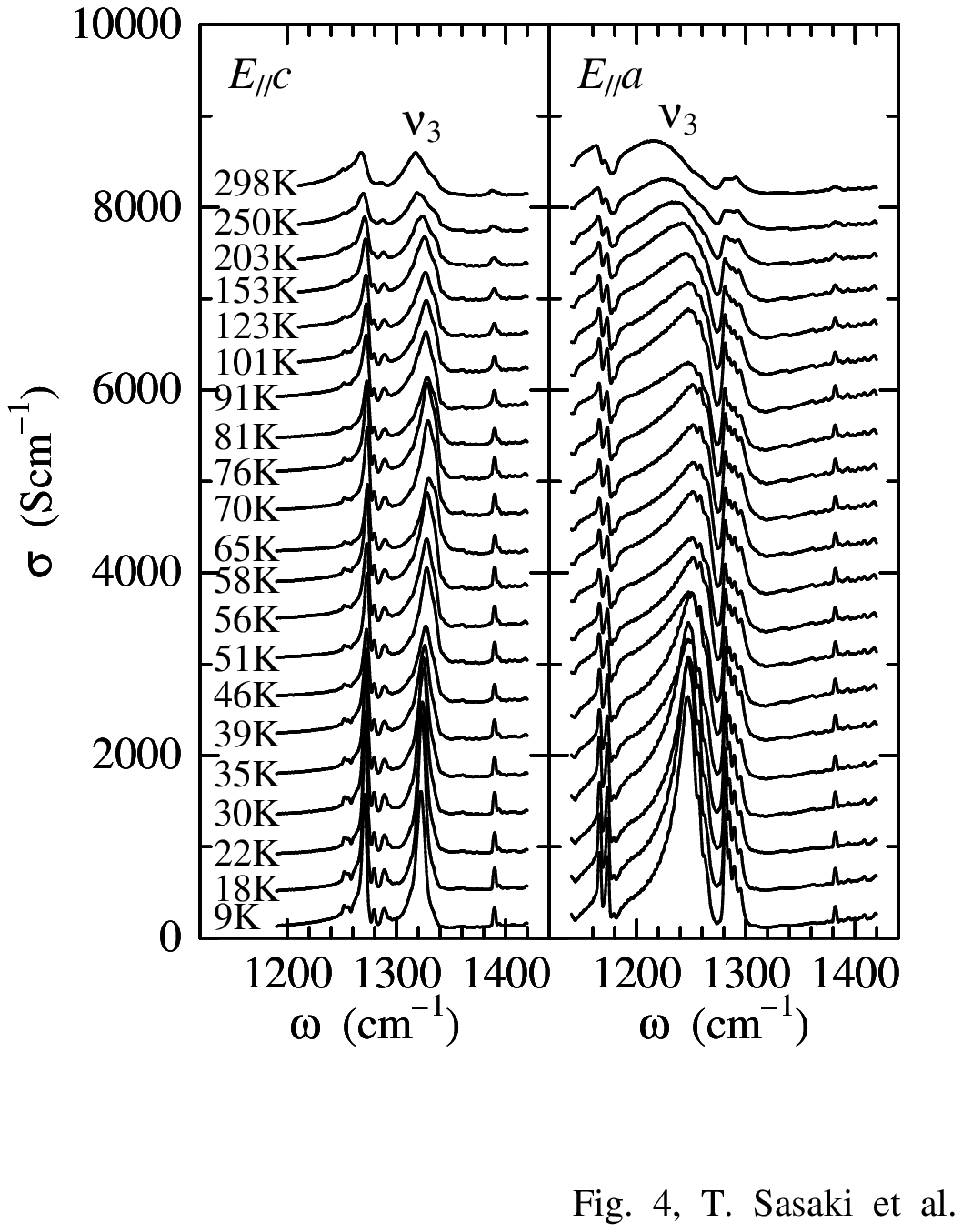}
\caption{The conductivity spectra near 1300 cm$^{-1}$ of $\kappa$-(BEDT-TTF)$_{2}$Cu[N(CN)$_{2}$]Cl for $E \parallel$ $c$-axis (left) and $E \parallel$ $a$-axis (right).  The spectra are plotted with shifting the conductivity by 400 Scm$^{-1}$ each. }
\end{figure}

The change of the electronic states in these salts is reflected in the totally symmetric internal vibration $\nu_{3}(a_{g})$ mode of the BEDT-TTF molecule through the EMV coupling.  
Figures 3 and 4 show the optical conductivity in the frequency range near 1300 cm$^{-1}$ of $\kappa$-(BEDT-TTF)$_{2}$Cu[N(CN)$_{2}$]Br and $\kappa$-(BEDT-TTF)$_{2}$Cu[N(CN)$_{2}$]Cl, respectively.  
The plotted conductivity value of these spectra are shifted from the data at the lowest temperature by 150 and 400 Scm$^{-1}$ each for $\kappa$-(BEDT-TTF)$_{2}$Cu[N(CN)$_{2}$]Br and $\kappa$-(BEDT-TTF)$_{2}$Cu[N(CN)$_{2}$]Cl, respectively.   
The sharp peak structures in the spectra have been assigned to the vibrational modes as follows.\cite{Eldridge1,Kornelsen1}
The peaks around 1170 cm$^{-1}$ and 1270 - 1290 cm$^{-1}$ have been assigned to $\nu_{67}$ and $\nu_{5}$ modes, both of which involve the CH$_{2}$ motion, respectively.  
The peak at 1380 - 1390 cm$^{-1}$ has been assigned to $\nu_{45}$.
These peaks do not change the frequency with temperature so much.  
On the other hand, the peaks at 1320-1330 for $E \parallel$ $c$-axis and 1250-1270 cm$^{-1}$ for $E \parallel$ $a$-axis shift to higher frequency below about 30-40~K in $\kappa$-(BEDT-TTF)$_{2}$Cu[N(CN)$_{2}$]Br, and to lower frequency below 50-60~K in $\kappa$-(BEDT-TTF)$_{2}$Cu[N(CN)$_{2}$]Cl. 
The former temperature corresponds to $T^{*}$ and the latter $T_{\rm ins}$.  
These relatively broad peaks have been assigned to the $\nu_{3}(a_{g})$ mode. 
The opposite trend of the temperature dependence of the peak positions has been reported on both salts.\cite{Eldridge1,Kornelsen1,Griesshaber}
The present experiments show clearly in the first time that the shift of the $\nu_{3}(a_{g})$ mode is not monotonic with temperature but correlated to $T^{*}$ in $\kappa$-(BEDT-TTF)$_{2}$Cu[N(CN)$_{2}$]Br, and $T_{\rm ins}$ and $T_{N}$ in $\kappa$-(BEDT-TTF)$_{2}$Cu[N(CN)$_{2}$]Cl.  
In concurrence with such frequency shifts, the shape of the conductivity peaks change at $T^{*}$ and $T_{\rm ins}$.  
The peak becomes sharper and larger below $T_{\rm ins}$ in $\kappa$-(BEDT-TTF)$_{2}$Cu[N(CN)$_{2}$]Cl, while the further broadening seems to be observed below $T^{*}$ in $\kappa$-(BEDT-TTF)$_{2}$Cu[N(CN)$_{2}$]Br.  
But the latter broadening can not be concluded because of overlapping with the Fano-type antiresonance of the CH$_{2}$ vibrations.  

\begin{figure}
\includegraphics[viewport=1cm 2.5cm 20cm 27cm,clip,width=0.9\linewidth]{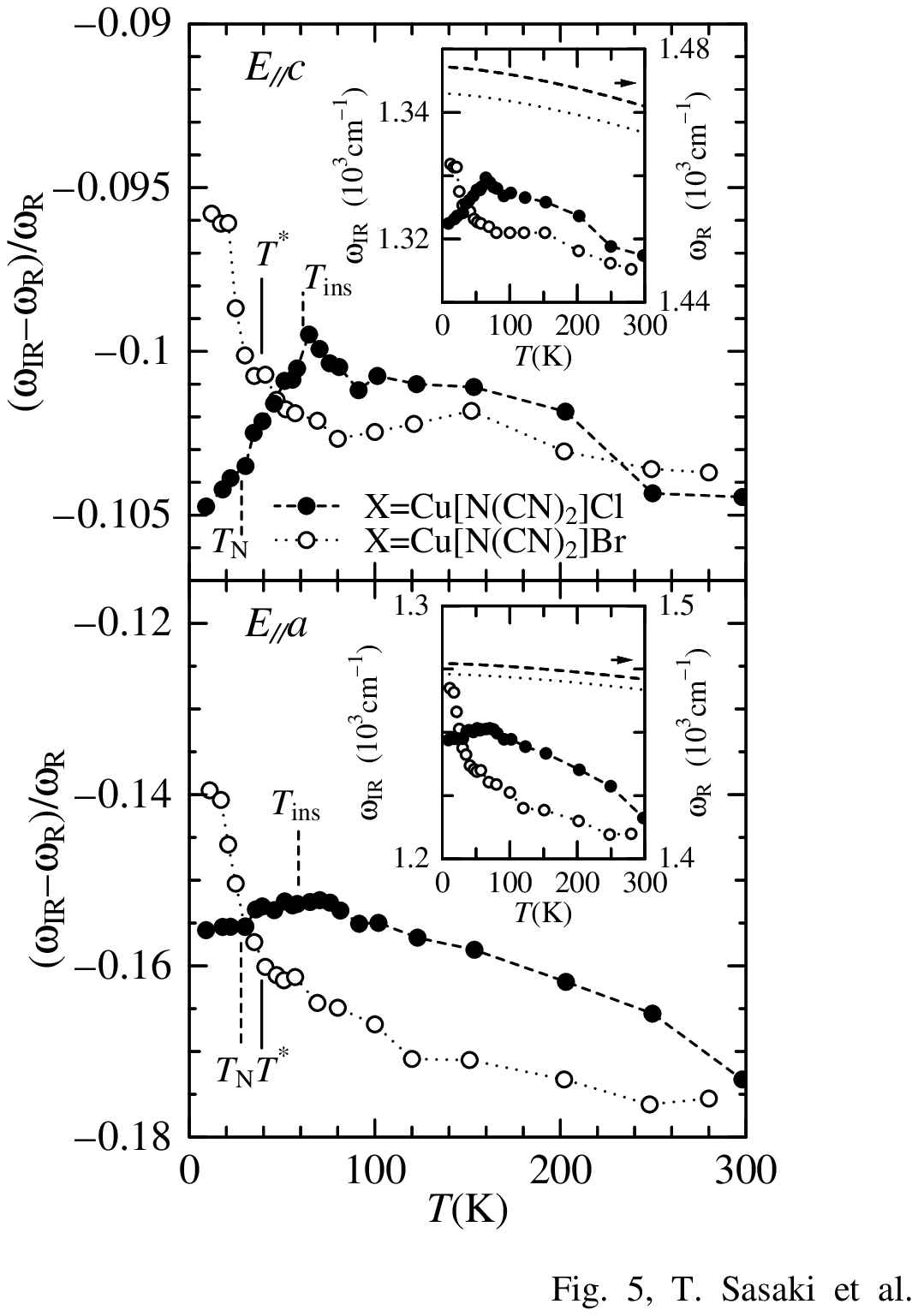}
\caption{Temperature dependence of the IR conductivity peak frequency ($\omega_{\rm IR}$) of the $\nu_{3}(a_{g})$ mode normalized by the same mode of the Raman shift ($\omega_{\rm R}$) in $\kappa$-(BEDT-TTF)$_{2}$Cu[N(CN)$_{2}$]Br (open circles and dotted line) and $\kappa$-(BEDT-TTF)$_{2}$Cu[N(CN)$_{2}$]Cl (filled circles and broken line) for $E \parallel$ $c$-axis (upper panel) and $E \parallel$ $a$-axis (lower panel).  Insets show the temperature dependence of $\omega_{\rm IR}$ and $\omega_{\rm R}$.}
\end{figure}

The peak frequency $\omega_{\rm IR}$ of the $\nu_{3}(a_{g})$ mode is shown in the inset of Figure 5.  
The peak frequencies in both salts and both polarization directions increase monotonically with decreasing temperature from room temperature.  
Below about $T_{\rm ins} \simeq$ 50-60~K, however, the peak frequency starts to shift to lower frequency in both polarization directions of $\kappa$-(BEDT-TTF)$_{2}$Cu[N(CN)$_{2}$]Cl.  
Then a small downward jump ( $\sim$ 1.5 cm$^{-1}$ for $E \parallel c$-axis and $E \parallel a$-axis) appears at $T_{N} \simeq$ 28~K.  
The magnitude of the jump is rather small in comparison with the entire temperature dependence.  
To the contrary, the same $\nu_{3}(a_{g})$ mode of superconducting $\kappa$-(BEDT-TTF)$_{2}$Cu[N(CN)$_{2}$]Br shifts to the higher frequency abruptly below $T^{*} \simeq$ 38~K in both the polarization directions.  
The magnitude of the shift below $T^{*}$ is rather large in comparison with the change at $T_{\rm N}$ in $\kappa$-(BEDT-TTF)$_{2}$Cu[N(CN)$_{2}$]Cl.
The change of the frequency at $T_{\rm c}$ is not known because the lowest temperature measured does not reach to $T_{\rm c}$, although the large frequency changes were observed in the transverse acoustic phonon mode below $T_{\rm c}$.\cite{Pintschovius}   

The characteristic temperature dependence of the $\nu_{3}(a_{g})$ mode in the IR optical conductivity can be attributed to the change of the electronic states through the EMV coupling, because the same vibrational mode measured by the Raman shift experiments shows only the monotonic temperature dependence of the frequency $\omega_{\rm R}$ from room temperature to 10~K.\cite{Eldridge2,Maksimuk}  
The main panel of Fig. 5 shows the temperature dependence of the relative frequency shift of the $\nu_{3}(a_{g})$ mode in the IR spectra, which is normalized by $\omega_{\rm R}$ shown in the inset.
The IR frequency $\omega_{\rm IR}$ is expected to be smaller than the bare phonon frequency  $\omega_{\rm 0}$ ($\simeq \omega_{\rm R}$) when the transfer integral $t_{\rm dimer}$ for hopping between molecules within the dimer is larger than $\omega_{0} (\simeq \omega_{\rm R}$).\cite{Jacobsen,Rice,Rice2}
These plots support further that the anomalous frequency shifts of the $\nu_{3}(a_{g})$ mode at $T^{*}$  in $\kappa$-(BEDT-TTF)$_{2}$Cu[N(CN)$_{2}$]Br, and $T_{\rm ins}$ and $T_{N}$ in $\kappa$-(BEDT-TTF)$_{2}$Cu[N(CN)$_{2}$]Cl are caused by changing not the molecular vibration itself but the electronic states on the dimers through the EMV coupling.  

In view of such change of $\omega_{\rm IR}$ of the $\nu_{3}(a_{g})$ mode, let us then consider the relation to the electronic states.  
In the $\kappa$-type BEDT-TTF system, the effective half filling bands are expected and results in the Mott insulator due to the strong dimerization.\cite{Kanoda1,Kanoda2,McKenzie2} 
This situation is realized below $T_{\rm ins}$ in $\kappa$-(BEDT-TTF)$_{2}$Cu[N(CN)$_{2}$]Cl.
Large negative frequency shift from the bare phonon frequency observed below $T_{\rm ins}$ corresponds to the enhancement of the EMV coupling.  
The large EMV coupling on the dimer system implies that the carriers tend to localize at the dimer and then the conductivity becomes small.  
The larger and sharper peak shape of the $\nu_{3}(a_{g})$ mode also demonstrates the localization of a carrier (one hole) at each dimer.  
Then the system has became the Mott insulating state before the AF static order appears at $T_{\rm N}$.  
In this context, the open of the charge gap has been reported below about $T_{\rm ins}$.\cite{Kornelsen1}
On the other hand, the positive frequency shift below $T^{*}$ indicates that the carriers on the dimer show the itinerant behaviour in the superconducting salt.  
This metallic behaviour is consistent with the results of the spin susceptibility which has shown that the density of states at Fermi level increases below $T^{*}$.\cite{Sasaki1}  
The resistivity also shows the quadratic temperature dependence below $T^{*}$, which suggests the Fermi liquid like metallic behaviour.\cite{Merino1}  
The superconductivity appears from such metallic state.

A possibility of the fluctuation of the charge density wave or charge localization in the metallic state below $T^{*}$ has been proposed on the basis of the anisotropic behaviour of the resistivity and spin susceptibility.\cite{Sasaki1}  
A separation or distortion of the line shape of the $\nu_{3}(a_{g})$ mode is expected in such instances.
In fact the clear peak separation has been observed in the charge ordered insulating state of the $\theta$-type BEDT-TTF system,\cite{Yamamoto} where the separation width is related to the degree of the charge disproportionation on each molecule.
The line shape of the $\nu_{3}(a_{g})$ mode in this study on the $\kappa$-type system, however, shows a sign of neither the separation nor the clear distortion below $T^{*}$, although the the shape becomes somewhat broadened.  
It needs further experiments and analysis for reaching the conclusion.

\begin{figure}
\includegraphics[viewport=1cm 2.5cm 20cm 27cm,clip,width=0.9\linewidth]{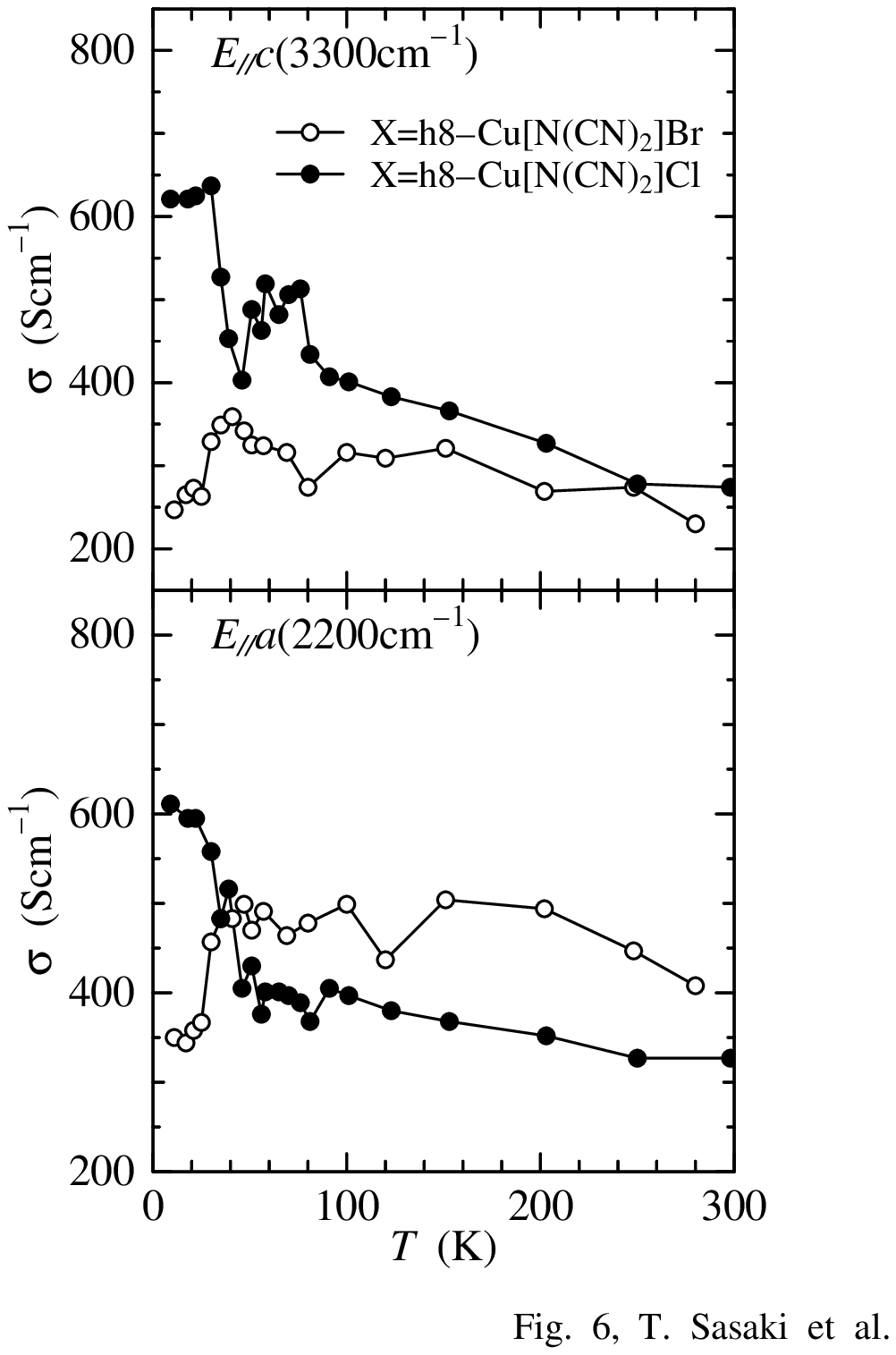}
\caption{Temperature dependence of the conductivity of $\kappa$-(BEDT-TTF)$_{2}$Cu[N(CN)$_{2}$]Br (open circle) and $\kappa$-(BEDT-TTF)$_{2}$Cu[N(CN)$_{2}$]Cl (filled circle) at 3300 cm$^{-1}$ for $E \parallel$ $c$-axis (upper panel) and 2200 cm$^{-1}$ for $E \parallel$ $a$-axis (lower panel).  Each frequency corresponds to the conductivity hump in the mid-IR region. }
\end{figure}

The EMV coupling reflects the intensity of the mid-IR band transition.  
The totally symmetric $\nu_{3}(a_{g})$ mode is not originally IR active.  
This mode can appear in the IR spectra by borrowing the intensity of the mid-IR band transition.@ 
Such mid-IR band transition should appear as a broad conductivity hump in the spectra.  
As has been reported on the several $\kappa$-type BEDT-TTF salts,\cite{Eldridge1,Kornelsen1,Tamura,Kornelsen2} a broad conductivity hump is observed in the mid-IR region which can be seen in Fig. 2.  
The characteristic feature is that the intensity of the hump changes with temperature but the peak frequency does not.  
The several possible origins of the hump have been proposed; the interband transition based on the tight-binding calculations\cite{Eldridge1,Kornelsen1,Tamura,Kornelsen2}, the correlated Mott-Hubbard bands,\cite{McKenzie2,Rozenberg} and the polaron absorption.\cite{Wang}  

Figure 6 shows the temperature dependence of the conductivity at the broad hump  frequency of 3300 cm$^{-1}$ for $E \parallel$ $c$-axis and 2200 cm$^{-1}$ for $E \parallel$ $a$-axis.  
One can find an obvious correlation between the temperature dependence of $\omega_{\rm IR}$ in Fig. 5 and the intensity of the mid-IR conductivity hump in both the superconducting and Mott insulating salts although the conductivity values are scattering a little.  
The intensity of the hump increases below $T_{\rm ins}$ and decreases below $T^{*}$ in $\kappa$-(BEDT-TTF)$_{2}$Cu[N(CN)$_{2}$]$Y$ with $Y = $ Cl and Br, respectively.  
The observation of such correlation, in other words, confirms that the origin of the hump is the mid-IR band transition.  
In addition the Mott-Hubbard picture should be considered rather than the tight-binding band model because the change of the band structure at $T_{\rm ins}$ and $T^{*}$ is difficult to be expected.
The model of the polaron formation\cite{Wang} also may have difficulty in explaining the almost temperature independence of the hump intensity between room temperature and $T^{*}$ or $T_{\rm ins}$ because the thermal excitation type of the polaron formation is expected.  
Further the almost constant of the spin susceptibility\cite{Sasaki1} and the Hall coefficient\cite{Murata} in relative broad temperature range from room temperature down to about 100~K is difficult to be explained by the polaron model.  

\begin{figure}
\includegraphics[viewport=1cm 6cm 20cm 24cm,clip,width=0.9\linewidth]{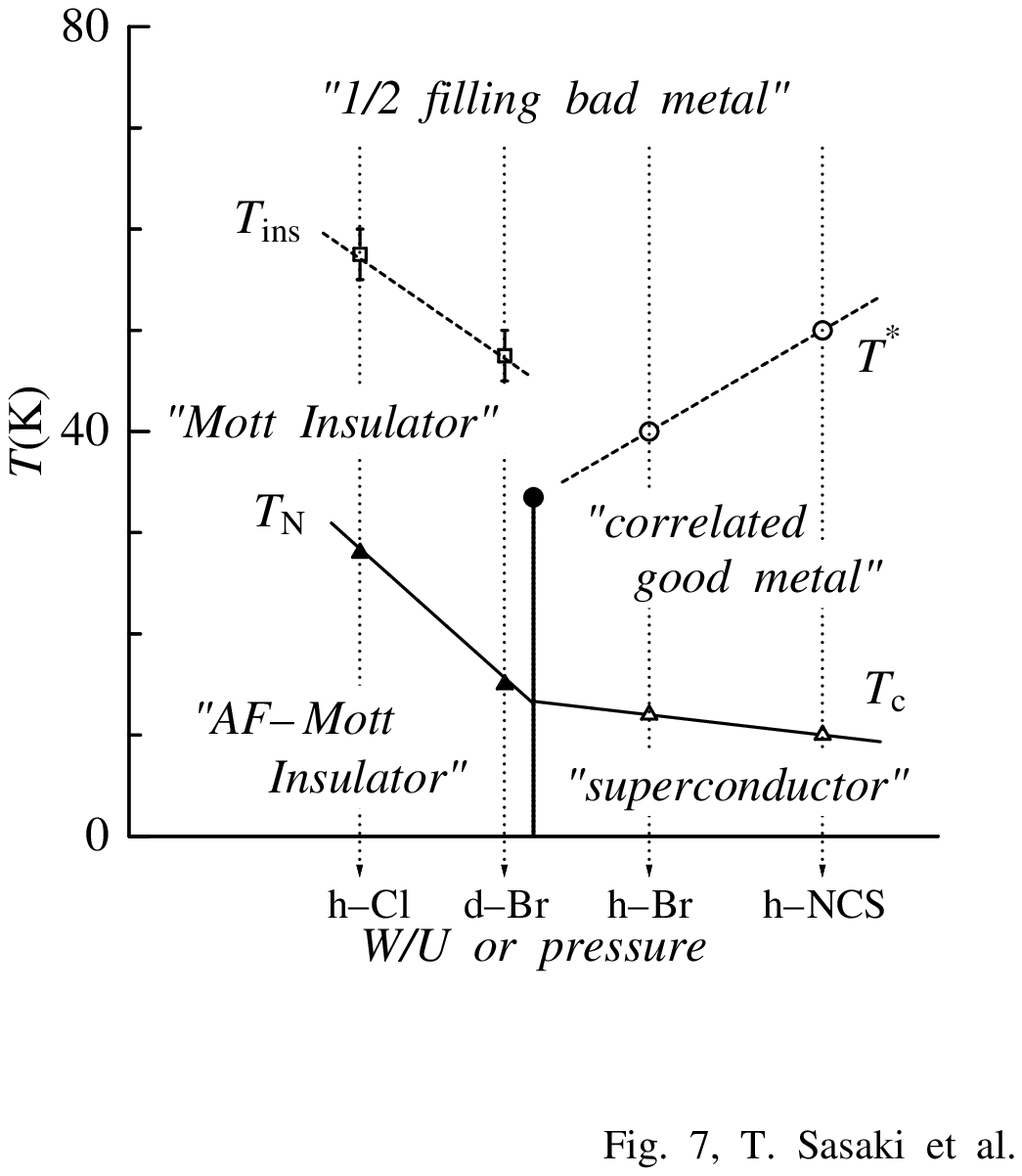}
\caption{Phase diagram of $\kappa$-(BEDT-TTF)$_{2}$$X$. The horizontal scale is arbitrary for the $W/U$ value of $\kappa$-(BEDT-TTF)$_{2}$$X$.  The h-Br, h-Cl and h-NCS denote the salts with $X$=Cu[N(CN)$_{2}$]Br, Cu[N(CN)$_{2}$]Cl and Cu(NCS)$_{2}$.  The d-Br represents the deuterated $X$=Cu[N(CN)$_{2}$]Br salt. Thick vertical line shows a first order Mott transition line.\cite{Lefebvre,Limelette,Kagawa} The filled circle indicates the critical end point of the transition. The dashed line terminated at the critical end point divides into two region at $T^{*}$,\cite{Sasaki1} the 1/2 filling bad metal at high temperature and the correlated good metal at low temperature.  The bad metal like as $X$=Cu[N(CN)$_{2}$]Cl with smaller $W/U$ than the critical value  results in the antiferromagnetic Mott insulator via the Mott insulator region.  In contrast, the bad metal with larger $W/U$ value can turn out to be the superconductor without the first order transition in only the case of passing through the correlated good metal region.}
\end{figure}

We shall discuss the behavior of the mid-IR conductivity hump and the resulting EMV coupling in view of the Mott-Hubbard picture.
The anisotropic triangular dimer Hubbard model has been proposed by Kino and Fukuyama\cite{Kino} for describing the electronic states of $\kappa$-type BEDT-TTF system.  
The model suggests the importance of the correlation between the effective on-site Coulomb repulsion $U$ on a single dimer and the interdimer hopping integrals $t$'s. 
Many of the essential features are well described and the critical $U_{c}$ value for the Mott insulating transition has been estimated to be $U_{c}/4t \sim U_{c}/W \sim 1$ in the Hartree-Fock approximation.  
The $\kappa$-type BEDT-TTF salts may be close to the transition because of $U \sim 2t_{\rm dimer} \sim$ 0.4 eV and $t \sim$ 0.1 eV, where $t_{\rm dimer}$ is the intradimer hopping integral.\cite{Kanoda1,Kanoda2,McKenzie2}  
In the optical conductivity in such Mott-Hubbard system, the broad mid-IR band transition between the upper and lower Hubbard bands is expected to appear around $\omega \sim U$ in the insulator region. 
The observation of the enhancement of the mid-IR conductivity hump in $\kappa$-(BEDT-TTF)$_{2}$Cu[N(CN)$_{2}$]Cl below $T_{\rm ins}$ can be understood in this picture and may confirm that the insulating state results in the Mott insulator.  

In the metallic $\kappa$-(BEDT-TTF)$_{2}$Cu[N(CN)$_{2}$]Br, the position of the mid-IR conductivity hump is the same with that in $\kappa$-(BEDT-TTF)$_{2}$Cu[N(CN)$_{2}$]Cl, and it does not change at $T^{*}$.  
This demonstrates that the mid-IR band transition between the upper and lower Hubbard bands keeps a certain amount of the intensity even below $T^{*}$ without changing the energy.  
The dynamical mean-field theory (DMFT)\cite{Rozenberg,Georges,Merino1} for calculating the Hubbard model predicts that a quasiparticle peak grows at the chemical potential in the metallic state below a temperature $T_{0}$ which is much smaller than $U$ and $W$.  
For temperature smaller than $T_{0}$ a Drude-like response appears at $\omega \sim$ 0 and the new peak forms at $\omega \sim U/2$, which corresponds to the transition between the coherent quasiparticle band and the upper and lower Hubbard bands, while a broad peak forms at $\omega \sim U$ in the insulating state above $T_{0}$.\cite{McKenzie2,Rozenberg}  
It is interesting to compare the metallic state found in $\kappa$-(BEDT-TTF)$_{2}$Cu[N(CN)$_{2}$]Br below $T^{*}$ to the coherent metal state expected below $T_{0}$ in DMFT.  
The expected new broad band, however, can not be found at $\omega \sim U/2$ in the conductivity spectra below $T^{*}$.  
The spectrum weight of the broad hump above $T^{*}$ moves not to the $\omega \sim U/2$ position but to the Drude response at $\omega \sim 0$ at low temperature.  
The DMFT calculations have shown also that the large phonon frequency shift takes place at $T_{0}$ in particular phonons at $\omega \sim U/2$.\cite{Merino2}  
The Raman frequency $\omega_{\rm R}$ of $\nu_{3}(a_{g})$ near $\omega \sim U/2$, however, does not change at $T^{*}$, while the IR conductivity frequency $\omega_{\rm IR}$ changes dramatically in line with the intensity of mid-IR conductivity hump at $\omega \sim U$.  
It follows from these points that the metallic state below $T^{*}$ still holds a feature of the Mott-Hubbard picture with a certain amount of the state density at $\omega \sim 0$.  
Then the state can be called as the {\it correlated good metal}.
But the intensity of the coherent quasiparticle peak may not be so strong enough to realize the transitions from the quasiparticle band to the upper Hubbard band and from the lower Hubbard band to the quasiparticle band.
Therefore it is not clear at present whether the observed $T^{*}$ corresponds to the coherent temperature $T_{0}$ in DMFT.  
It is noted that the second order phase transition at $T^{*}$\cite{Lang2,Mueller} has been proposed while DMFT shows that $T_{0}$ is a crossover temperature.\cite{McKenzie2} 

\section{Conclusion}

We have measured the optical conductivity in the IR region of the Mott system $\kappa$-(BEDT-TTF)$_{2}$$X$.  
The specific molecular vibration mode $\nu_{3}(a_{g})$ of the BEDT-TTF molecule shows the sudden frequency shift at $T_{\rm ins}$ and $T^{*}$ where the system changes to the Mott insulator and the {\it correlated good metal}, respectively.  
These frequency shifts have strong correlation with the intensity of the mid-IR conductivity hump.  
The mid-IR conductivity hump is explained by the transition between the upper and lower Mott-Hubbard bands. 
The characteristic temperatures $T_{\rm ins}, T_{\rm N}, T_{\rm c}$ and $T^{*}$ are summarized as the schematic phase diagram of $\kappa$-(BEDT-TTF)$_{2}$$X$ in Fig. 7.  
The results of the similar optical conductivity experiments in the $\kappa$-(BEDT-TTF)$_{2}$Cu(NCS)$_{2}$ and the deuterated $\kappa$-(BEDT-TTF)$_{2}$Cu[N(CN)$_{2}$]Br are also plotted in the figure.\cite{Sasaki2}  
Above $T_{\rm ins}$ and $T^{*}$ the system has almost the same electronic state of the so-called {\it half filling bad metal},\cite{McKenzie2,Merino1,Emery}  where the transitions between the upper and lower Mott-Hubbard bands are dominant. 
In the small $W/U$ side, the bad metal changes to the {\it Mott insulator} at $T_{\rm ins}$ and then to the {\it AF-Mott insulator} at $T_{\rm N}$.  
On the other hand, the half filling bad metal in the large $W/U$ side is altered to the {\it correlated good metal} at $T^{*}$.  
The possible fluctuation of a density wave formation\cite{Sasaki1} is not conclusive in the metallic state at present.
The superconducting state can be realized only from the good metal by the second order phase transition.  

The Mott insulator and the good metal at high temperature and the AF-Mott insulator and the superconductor are separated by the contiguous first order transition line.\cite{Lefebvre,Limelette,Kagawa}  
The first order phase transition line is terminated at the critical end point.  
The recent observation of the phase separation\cite{Miyagawa2} and the field induced phase alternation\cite{Taniguchi} in the system being situated next to the first order transition may be explained by the spacial inhomogeneity consisting of the Mott insulator and the good metal or the AF-Mott insulator and the superconductor.  
The possible origin of the spacial inhomogeneity may have the close relation to the ethylene disorder,\cite{Singleton2,Yoneyama} that is, the glass transition\cite{Mueller} due to the freezing of the ethylene motion at higher temperature in addition to the competition of the free energy of each state.   

The horizontal axis of the phase diagram can convert to the pressure through $W$.  
The strong pressure dependence of the $\nu_{3}(a_{g})$ mode of $\kappa$-(BEDT-TTF)$_{2}$Cu(NCS)$_{2}$ at room temperature has been reported.\cite{Klehe}  
The pressure dependence in the half filling bad metal can be also understood in this picture.  
The increasing of $W$ leads the bad metal to the normal metal with less electron correlation and then the EMV coupling becomes weak.  
In fact the mid-IR conductivity hump is  suppressed under pressure.\cite{Klehe}  
It must be interesting to study the optical properties near the first order transition under pressure or in the materials controlled with the chemical pressure by way of the partial substitution of the molecular elements.\cite{Kawamoto2}
 
\begin{acknowledgments}
We thank R.~H.~McKenzie, J.~Merino and M.~Lang for helpful discussions.
A part of this work was performed in the Spectroscopy laboratory, the Material Design and Characterization Laboratory, ISSP, University of Tokyo.  
This work was partly supported by a Grant-in-Aid for Scientific Research (C) (No. 15540329) from the Ministry of Education, Science, Sports, and Culture of Japan.  

\end{acknowledgments}


\end{document}